\documentclass[preprint]{aastex}
\usepackage{natbib}
\usepackage{graphicx}
\usepackage{wasysym}
\usepackage{txfonts}
\usepackage{ulem}
\def\fH2{\mbox {f$_{{\rm H}_2}$}}

\def\EBV{\mbox{\rm{E(B-V)}}}

\def\AV{\mbox{A$_{\rm V}$}}

\def\nH2{\mbox{${\rm n}(\HH$)}}
\def\enH2{\mbox{$n_{(\HH$)}}}

\def\pccc{~{\rm cm}^{-3}} 
 
\def\pcc {\mbox{${~{\rm cm}^{-2}}$}}

\def\Tsub#1 {\mbox{${\rm T}_{\rm #1}$}}
\def\TK  {\Tsub K }

\def\Tsp {\Tsub sp }

 \def\arcmin{\mbox{$^{\prime}$}}

\def\degr{\mbox{$^{\rm o}$}}
\def\p{\mbox{$^+$}}

\def\h13cop{\mbox{{H$^{13}$CO\p}}}

\def\c3h2{\mbox{C$_3$H$_2$}}

 \def\R0{R$_0$}
\def\G0{\mbox{G$_0$}}

\def\ddeg{{}^\circ\kern-.1em}

\def\kms{\mbox{km\,s$^{-1}$}}

\def\m1{\mbox{$^{-1}$}}

\def\E#1 {$10^{#1}$}
\def\E#1 {E{#1}}
\def\P#1,{$\nH2\TK~=~#1\times~10^4\pccc$~K}
\def\ec#1,#2,#3,{#1\,(#2)\E{#3}}

\def\H3{\mbox{H$_3$}}
\def\Lya{\mbox{Ly-$\alpha$}}

\def\RH2{\mbox{R$_{\rm G}$}}
\def\g13{\mbox{g$_{13}$}}

\def\kHeH2{\mbox{$k_{ He-\HH}$}}

\def\tim#1,#2{\mbox{{$#1\times10^{#2}$}}}

\def\WHI{\mbox{$\Upsilon_{{\rm HI}}$}}

\def\l21{\mbox{{$\lambda$21cm}}}

\newcommand{\emm}[1]{\ensuremath{#1}}   % ensures math mode.
\newcommand{\emr}[1]{\emm{\mathrm{#1}}} % uses math roman fonts.
\newcommand{\hcop}{\emr{HCO^+}} 
\newcommand{\HI}{\emr{HI}} 
\newcommand{\HH}{\emr{H_2}}

\slugcomment{generated \today}

\shorttitle{AGN: use \WHI}
\shortauthors{Harvey Liszt}

\begin{document}


%\title{\EBV\ and $\tau({\rm HI})$ }
\title{Reconciling X-ray and \l21\ absorption gas column densities toward obscured AGN}


\author{Harvey Liszt}
\affil{National Radio Astronomy Observatory \\
        520 Edgemont Road, Charlottesville, VA 22903}

\email{hliszt@nrao.edu}

%\today



\begin{abstract}
Hydrogen column densities inferred from X-ray absorption are typically 5 - 30 
times larger than the neutral atomic hydrogen column densities  derived 
from \l21\ HI absorption toward radio-loud active galactic nuclei.  Some 
part of the difference 
is ascribed to uncertainty in the spin temperature \Tsp\ = 100 K that 
is often  used to convert \l21\ HI absorption to N(HI).  Here we propose another 
way to  infer the gas column from HI absorption. In our Galaxy there is a nearly 
linear correlation between the inteferometrically-measured integrated \l21\ 
absorption \WHI\ and reddening, \WHI\ $\propto$ \EBV$^{1.10}$ for 
\WHI\ $\ga 0.7$ \kms\ or \EBV\ $\ga 0.04$ mag.
Scaling \EBV\ then provides the total gas column density N(H) from the same 
dust column that is responsible for optical obscuration and X-ray absorption, 
without calculating N(HI). Values of N(H) so derived typically exceed N(HI) 
by a factor 4 because the ubiquitous Galactic \l21\ HI absorption samples 
only a portion  of the interstellar gas.  If the well-studied case of Hydra-A 
is a guide, even very large disparities in X-ray and \l21\ gas  column
densities can be explained by resolving the core radiocontinuum and 
inferring N(H) from \l21\ absorption. Milky Way conditions
are often invoked in discussion of obscured AGN, so the empirical
relationship seen in the Milky Way should be a relevant benchmark.

 \end{abstract}



%s1
\section{Introduction}

It is a striking realization of contemporary astrophysics that active 
galactic nuclei (AGN) could be optically obscured and radio quiet, effectively 
masking some of the most energetic phenomena in the Universe \citep{HicAle+18}. 
Even when the nuclear region is visible, obscuration can affect its type
classification \citep{MalBas+20}.  The intervening gas column density needed 
to obscure and/or misclassify an AGN, log(N(H)$\pcc$ = 21.6 - 22.0 
\citep{PanBas+16,UrsBas+20} can accrue across a few kpc in the distributed 
material in the disk of a spiral galaxy, or across a single translucent molecular 
cloud.

To determine the column density of the obscuring material, the metal content in 
the obscuring dust column may be inferred from soft X-ray absorption, and neutral 
atomic hydrogen column densities N(HI) are occasionally measured in 
\l21\ HI absorption toward radio-loud AGN. When absorption is detected in both tracers, 
high-resolution maps of the HI absorption provide useful constraints on the kinematics
and distribution of the obscuring material \citep{Tay96}. But the total hydrogen 
column densities inferred from X-ray absorption are always appreciably larger than 
N(HI) \citep{Tay96,OstMor+16,MosAll+17,GloAll+17,MorOos18,GloAll+19,SadMos+20}, 
casting doubt on the relevance of the comparison of the HI and X-ray results.  

Significant disparity may arise from failure to resolve or otherwise isolate 
the radio continuum contribution of the core, as seen from comparing the VLA 
results of \cite{DwaOwe+95} at arcsecond resolution toward Hydra-A and the 
better-resolved VLBA observations of \cite{Tay96}. \cite{Tay96} observed much 
larger N(HI) than was detected earlier and found much larger N(HI) toward the 
core compared with the adjacent radio jets. This removed all but a factor 
$\approx$ 3 of the difference in column densities toward the core, which is 
explained here.   

AGN-obscuring material is also observed in mm-wave absorption from small trace 
molecules whose column densities, when inter-compared, show clear similarities to 
relative molecular abundances  seen in gas in the disk and low halo of the Milky 
Way \citep{RosEdg+19,RosEdg+20}. This circum-AGN chemistry might be used to 
derive improved knowledge of physical properties in the gas if absolute 
molecular abundances relative to hydrogen can be determined. 

Deriving the total gas column density from \l21\ HI absorption and understanding 
and alleviating the disparity between N(HI) derived from \l21\ HI absorption and 
N(H) derived in soft X-ray absorption constitute the subject of this work. In Section 2 
we discuss the interpretation of \l21\ HI absorption in the Milky Way in the context
of its correlation with reddening.  We discuss how the relationship between \WHI\ and \EBV\ 
provides a direct means of deriving N(H) without the need to derive 
N(HI), although knowledge of N(HI) may be  useful for inferring the character of 
the intervening gas. We show that the empirical relationship between \WHI\ and \EBV\
in the Milky Way can usefully be applied to resolve an outstanding discrepancy
in gas column densities toward Hydra-A and we suggest that it should be a
useful benchmark when other aspects of conditions in the Milky Way are invoked
to discuss obscured AGN. Section 3 is a summary.

\section{HI absorption, \WHI\ and \EBV}

\subsection {\WHI\ and N(HI)}

For a given integrated \l21\ HI optical depth 
\WHI\ $\equiv \int\tau({\rm \HI}) ~{\rm dv}$
measured in units of \kms\ the inferred HI column density is 
N(HI) $= 1.823\times 10^{18}~\pcc~$\Tsp\ \WHI\ 
\footnote{In terms of a Gaussian profile with peak optical depth $\tau_0$ 
and full width at half maximum $\Delta$V, \WHI\ = 1.064 $\tau_0 ~\Delta$V.}.  
where \Tsp, the so-called spin temperature, is the excitation temperature of the 
\l21\ transition. The \l21\ HI optical depth is inversely proportional to \Tsp.
 \Tsp\ is expected 
to equal the ambient kinetic temperature in strongly-absorbing cold neutral medium 
(CNM) at \TK\ $\la 100$K but may fall below the kinetic temperature in 
weakly-absorbing warm partially-ionized gas at 8000 K, largely depending on
the \Lya\ radiation field \citep{Lis01,SeoKim20}.

Toward AGN, HI column densities are almost always inferred by assuming \Tsp\ = 100K
\citep{Tay96,OstMor+16,MosAll+17,GloAll+17,MorOos18,GloAll+19,SadMos+20}, that is very 
broadly characteristic of the CNM providing most of the \l21\ absorption 
in the Milky Way. The HI column densities derived in this way are invariably 
smaller than the absorbing column densities that are inferred from X-ray 
absorption, even toward the same AGN. Amounts range from factors of a few to
more than thirty. In at least one case \citep{DwaOwe+95,Tay96} most of the disparity
is explained by failure to resolve the radio core, leaving unaccounted only 
a factor of a few.

In fact this is broadly in line with contemporary consideration of the interstellar 
medium (ISM) in the Milky Way, which recognizes that the ubiquity of \l21\ absorption 
occurs despite the fact that its main carrier, the CNM, is only one component 
of the multi-phase medium. The SPONGE HI absorption survey investigators 
\citep{SPONGE18} concluded that 50\% of the atomic hydrogen was detected in \l21\ 
absorption. Estimates of the molecular fraction in 
the local diffuse ISM are in the range 25\% - 40\% \citep{BohSav+78,LisLuc02,LisPet+10}, 
or roughly one-third  , with a smaller contribution, of order one-sixth, from ionized gas.  
Thus about half of the local gas  is in atomic form and half of that, one-quarter, 
is sampled in \l21\  absorption.  

The point is that Milky Way observers recognize that 
there should be substantial 
differences between N(H) and N(HI) derived from \l21\ HI absorption, even in the
predominantly atomic diffuse ISM, and when the best value of \Tsp\ is employed.

\subsection{\WHI, \EBV\ and N(H)}

Figure 1 shows that there is an empirical correlation between the equivalent \EBV\ 
derived from far-IR dust emission by \cite{SchFin+98} and interferometric measurements of 
\WHI\ \citep{Lis19}.  The interferometric \WHI\ data set includes 160 measurements
from diverse sources including the Arecibo (RIP) survey of 
\cite{DicKul+83} and the SPONGE survey from the VLA \citep{SPONGE18}, along with the smaller datasets of 
\cite{KanBra+11} and \cite{RoyFra+17} from the VLA and GMRT and a handful of VLA measurements taken 
to complement mapping around compact mm-wave continuum sources as reported in \cite{LisPet+10} and 
\cite{LisPet12}.

For the data  with \WHI\ $>$ 0.7 \kms\ and \EBV\ $\ga 0.04$ mag

$$ \WHI\ =
  (13.8\pm0.7)~\EBV^{(1.10\pm0.03)}~\kms \eqno(1) $$.

\l21\ absorption is less reliably detected for \EBV\ $\la 0.04$ mag or \WHI\ $< 0.7$ 
\kms. As \HH\ becomes abundant only for \EBV\ $> 0.07$ mag  \citep{BohSav+78,SavDra+77}, the 
explanation must reside with the atomic gas:  Apparently, less-absorbent warm neutral 
or ionized gas makes up a larger fraction of the hydrogen. \cite{KanBra+11} were the 
first to note this phenomenon in a general way but it actually sets in at 
\EBV\ $\la 0.04$ mag that is asociated with a 50\% higher column density, 
$3\times10^{20}\pcc$, than \cite{KanBra+11} discussed. Surveys of \l21\ absorption toward 
AGN can not be expected to trace obscuring gas down to arbitrarily small column 
densities.

We ignored the datapoints in Figure 1 at \EBV\ = 8-14 mag, representing 
extinctions comparable to or higher than those toward Sgr A$^*$ at the Galactic center, 
 which lie about 0.3 dex below the regression line.
Such \EBV\ are  well above the expected range of validity of the  
reddening derived from FIR dust emission, according to \cite{SchFin+98}.

The utility of the \WHI-\EBV\ relationship in studies of obscured AGN is obvious.
It is unnecessary to assume an HI spin temperature and derive N(HI) when an estimate
of the gas associated with the obscuring dust column can be derived just by rescaling \WHI.  
As an example, \cite{RosEdg+20} recently detected many mm-wave molecular absorption 
lines toward the obscured AGN in Hydra-A and drew comparisons with comparable absorption 
line observations of CO, CN, HCN,\hcop, and SiO etc having similar column densities in 
observations of the diffuse ISM in the Milky Way.  They noted the map of \l21\ 
absorption by \cite{Tay96} who resolved the radio core and derived N(HI)
$= 1.4\times10^{22}\pcc$ in that direction with \Tsp\ = 100 K. \cite{RosEdg+20}
noted the X-ray absorption measurement N(H) $= 3.5\times10^{22}\pcc$ of
\cite{RusMcN+13}, representing a relatively small disparity between N(HI) and N(H)
as these things go, only a factor 3. But using Eqn 1 yields N(H) $= 4.1\times 10^{22}\pcc$
when used in conjunction with the appropriate value 
N(H)/\EBV\ = $8.3 \times 10^{21}\pcc$ {\rm H-nuclei} (mag)$^{-1}$ \citep{Lis14yEBV,Lis14xEBV}.

Other recent studies find or use ratios N(H)/\EBV\ =$7.7-9.4 \times 10^{21}
~{\rm H-nuclei} \pcc$ (mag)$^{-1}$ \citep{HenDra17,LenHen+17,LiTan+18} and \cite{HenDra20} use 
$8.8 \times 10^{21}~{\rm H-nuclei} \pcc$ (mag)$^{-1}$. Some of this variation is due to 
a suggested 14\%\ downward scaling \citep{SchFin11} of the all-sky reddening 
maps of \cite{SchFin+98} but the value we used is unaffected by this rescaling 
when used to derive N(H), for reasons discussed by \citep{Lis19}.

\subsection{Deriving N(H) from \WHI}

Figure 2 shows the data with the variables of Figure 1 interchanged
and taking N(H)/\EBV\ = $8.3\times 10^{21}~{\rm H-nuclei} \pcc$ mag$^{-1}$
on the scale at right.
% Note again that this scale is unaffected by a linear
%%rescaling of the reddening values of \cite{SchFin+98}  for the reasons given by 
%\cite{Lis19}, based on the methods used in earlier work.
The detections are well represented by two power laws expressed in Eqns 2a and 
2b, joined at \WHI\ = 0.72 \kms\ the lower limit that was used to fit the 
\EBV-\WHI\ relationship of Eq. 1. The upper limits taken from the work of 
\cite{RoyFra+17} do not present interesting constraints when the data are fit in this way.
A Monte Carlo analysis with Gaussian random errors in \WHI\ yielded
a 95\% confidence limit 0.18 dex \citep{Lis19}, or a factor 1.5, and the vertical 
bars along the power-law fits in Figure 2 are $\pm$0.18 dex, a factor 1.5.

In terms of the column density
$$ N({\rm H}) = A \WHI^{0.395}, \WHI\ \le 0.715~\kms \eqno(2a)$$
and 
$$ N({\rm H}) = B \WHI^{0.950}, 0.715~\kms \le \WHI \eqno(2b) $$
where A $ =5.43\times10^{20}\pcc$ and B $ = 6.54\times10^{20}\pcc$.  
The equivalent prediction for \EBV\ follows by dividing the predicted
column density by N(H)/\EBV.

The dashed gray line in Figure 2 gives N(HI) at \Tsp\ = 100 K when interpreted on
the column density scale at right in the Figure.  The \WHI\ values encountered in 
obscured AGN are  typically in the range \WHI\ $\approx$ 1-50 where the disparity 
between N(H) and N(HI) derived at \Tsp\ = 100 K is at least a factor 4.
In the Milky Way there are implied constraints on \Tsp\ as a function of \EBV\ 
and the appropriate value of \Tsp\ is below 100K for \EBV\ $\ga$ 0.8 mag,
at which point the optical depth correction to N(HI) derived by integrating
the HI emission profile in the optically thin limit is a factor 1.3 (see Figure 
4 of \cite{Lis19}). Thus the \Tsp\ values appropriate to the Milky Way are below 
100 K at the values of \WHI\ and \EBV\ that occur toward obscured AGN, widening 
the disparity between N(HI) and N(H). 

\subsection{Sightlines toward AGN with estimated \EBV\ and measured \l21\ absorption}

Table 1 summarizes information for five sources with \l21\ HI absorption measurements
and estimates of extinction or reddening based on optical spectra (ie not X-ray 
absorption). Shown are the Galactic foreground reddening, the intrinsic value in the 
AGN in the cited reference and the reddening predicted from Eqn 2a given \WHI\ and 
N(H)/\EBV\ = $8.3\times 10^{21}~{\rm H-nuclei} \pcc$ mag$^{-1}$.
Except toward 0414+534 the quoted values of the intrinsic reddening are 
taken from  published values of \AV\ using \EBV\ = \AV/3.1 as would 
be appropriate to the Milky Way. \cite{GloAll+19} derived \EBV\ toward 
1829-718 by fitting the observed optical spectrum to an unreddened AGN template.

The Galactic foreground reddening toward 0414+534 is high enough at its substantial 
Galactic latitude b= -31.03\degr\ to imply that this object lies behind  a diffuse 
molecular cloud in the Milky Way.  The foreground reddenings toward 0500+019 and 
1829-729 are also high enough to expect a substantial Galactic \HH\ column, athough 
probably not CO. 

The weak intrinsic \l21\ HI absorption toward 0414+534 is inconsistent 
with substantial reddening in a Milky Way environment and could result from the 
complex geometry of this lensed system, given that the \l21\ HI absorption was
synthesized  with an 0.8\arcmin\ $\times$ 8.5\arcmin\ beam. The moderate 
reddening toward 1829-718 is too small for the high \WHI\ that is measured there. 
Overall there is no trend toward a monotonic relationship between \WHI\ and 
\EBV\  inferred toward AGN.

%1
\begin{figure}
\includegraphics[height=7.7cm]{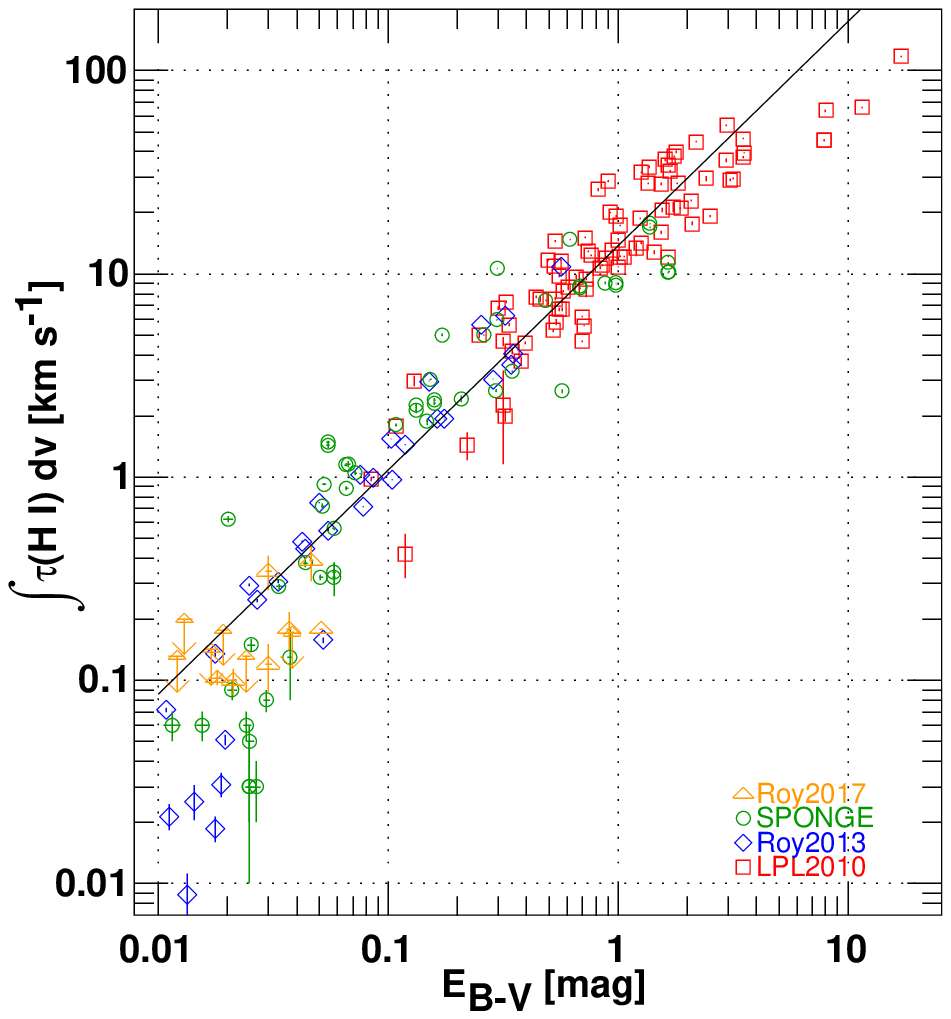}
 \caption[] {Integrated \l21\ \HI\ optical depth measured
interferometrically, plotted against IR dust emission-derived optical reddening 
from \cite{SchFin+98}. \HI\ optical depths are taken from \cite{RoyKan+13}
and \cite{RoyFra+17}, from the SPONGE survey \citep{SPONGE18} and from
the sample assembled by \cite{LisPet+10} (LPL), most of which was taken from
the survey of  \cite{DicKul+83}.  The regression fit for 160 sightlines with 
detections of HI absorption at \WHI\ $\ge 0.07$ \kms\ in the merged sample 
is given in Eqn 1.}
\end{figure}

%2
\begin{figure}
\includegraphics[height=7.7cm]{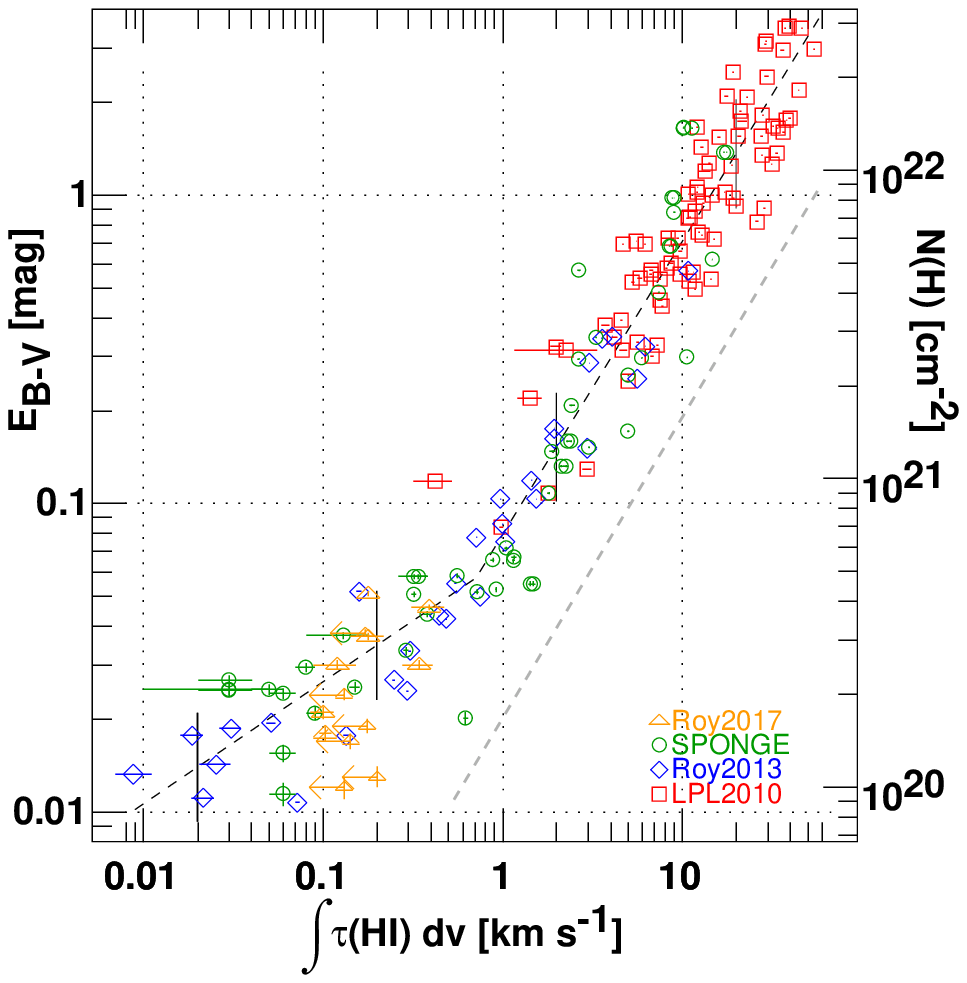}
 \caption[] {Reddening and implied hydrogen column density vs integrated
\l21\ optical depth.  Data plotted in Figure 1 are shown with the variables 
interchanged and fit with two power laws meeting at \WHI\ = 0.72 \kms\ aa 
given in Eqns 2a and 2b.  Shown along the fitted curves
are vertical markers denoting factor 1.5 ($\pm 1.76$ dex) variations.
The scaling to N(H) uses N(H)/\EBV\ = $8.3 \times 10^{21}\pcc$ mag$^{-1}$
and is invariant to a linear re-scaling of \EBV\ as discussed in Section 2.2.
The dashed grey curve shows N(HI) when \Tsp\ = 100K when read on the scale
at right.}
\end{figure}

%t1
\begin{table}
\caption{\WHI, foreground, observed and predicted \EBV }
{
\small
\begin{tabular}{lccccc}
\hline
AGN & \WHI\ & \EBV$_{\rm MW}$$^1$ &\EBV$_{\rm AGN}$$^2$ &\EBV$_{\rm pred}$ & References \\ 
    & \kms  & mag & mag & mag & \\       
\hline
0108+388 & 46.0 & 0.051 & $\ga 0.65$$^2$ & 3.00 & \cite{CarMen+98} \\   
0414+534 & 4.1  & 0.307 & $\approx 1.61$ & 0.29 & \cite{LawEls+95,MooCar+99} \\
0500+019 & 3.9  & 0.070 & $\ga 0.6$5 & 0.29 & \cite{CarMen+98} \\  
1504+377 & 24.4 & 0.014 & $\ga 2.2$ & 1.61 & \cite{CarMen+98} \\  
1829-718 & 27.4 & 0.081 & 0.455 & 1.83 & \cite{GloAll+19} \\
\hline
\end{tabular}
\\
$^1$from \cite{SchFin+98} \\
$^2$\EBV$_{\rm AGN}$ = estimated \AV/3.1 except toward 1829-718\\
\\
}
\end{table}

\section{Summary}

In the Milky Way the integrated \l21\ HI optical depth \WHI\ is very nearly linearly 
proportional to the optical reddening \EBV\ \citep{SchFin+98} when 
\WHI\ $\ga 0.7$ \kms\ and $\la$ 0.04 \EBV\ $\la$ 0.04 mag as shown in Figure 1 
and expressed in Eqn 1 \citep{Lis19}.  The existence of such
a relationship between quantities that are measured on angular scales of 6\arcmin\ 
in the FIR and  milli-arcseconds toward AGN (perhaps even obscured AGN) 
at \l21\ is quite remarkable.  Along more lightly-extincted sightlines at 
\EBV\ $\la 0.04$ mag and N(H) $\ga 3\times 10^{20}\pcc$  there is a higher 
proportion of warm atomic hydrogen and on average less \l21\ HI absorption 
per unit reddening.

The \WHI-\EBV\ relationship allows a direct derivation of the reddening and 
total column density of H-nuclei from observation of \WHI\ without the need to 
assume a spin temperature for the absorbing HI or to derive N(HI), although
the comparison between N(HI) and N(H) is informative on its own terms. Figure 2 
shows an inversion of the \WHI-\EBV\ relationship in Figure 1 to derive N(H) 
as a function of \WHI\ as expressed in Eqns 2a and 2b. For \WHI\ $> 0.7$ \kms\
one has N(H) $= 6.54\times10^{20}~{\rm H-nuclei}\pcc \WHI^{0.950}$ with a substantial
flattening at lower \WHI.  To derive these relationships we took
N(H)/\EBV\ = $8.3 \times 10^{21}~{\rm H-nuclei} \pcc$ mag$^{-1}$ that is 
appropriate to the methods used in our earlier work \citep{Lis14yEBV,Lis14xEBV}. 
Similar but not identical values N(H)/\EBV\ 
$= 7.7-9.4 \times 10^{21}~{\rm H-nuclei} \pcc$ mag$^{-1}$ have been used in
other recent studies (Section 2.2).  Some part of the variation in these values is due to
a suggested 14\% downward rescaling \citep{SchFin11} of the \EBV\ values of 
\cite{SchFin+98} but the column densities N(H) we calculate  are insensitive 
to such a scaling.

In Figure 2 it is seen that N(H)/N(HI) $\ga 4$ when N(HI) is derived from \l21\ HI 
absorption with \Tsp\ = 100 K as usually assumed in studies of obscured AGN.
Comparable or larger differences are seen in the Milky Way where it is understood
that the cold neutral medium providing the bulk of the \l21\ absorption represents 
only some 50\% of the atomic hydrogen overall \citep{SPONGE18} and the ISM is 
comprised of warm and cold, ionized and neutral, atomic and molecular gases.

Gas column densities derived from X-ray absorption toward obscured AGN are always 
much larger than N(HI) inferred from \l21\ HI absorption with a spin temperature 
\Tsp\ = 100 K, even in the same direction 
\citep{OstMor+16,MosAll+17,GloAll+17,MorOos18,GloAll+19,SadMos+20}. In the case of 
Hydra-A, resolving the radio core with the VLBA relieved all but a factor 3 of the 
disparity \citep{Tay96} and subsequent application of the \WHI-\EBV\ relationship 
yields closely equal N(H) $\approx 4\times10^{22} \pcc$ from both X-ray and \l21\ 
HI absorption measurements.  It may well be possible to relieve comparable disparities 
in other directions in the same way.

When conditions similar to those of the Milky Way are taken as relevant to observations 
of obscured AGN, the relationships between \WHI\ and \EBV\ or N(H) should be useful 
benchmarks.  

%The handful of sightlines toward obscured AGN 
%for which values of \WHI\ and \EBV\ (often lower limits) are given in the literature 
%(Table 1 and Section 3.1) do not show a consistent increase of \EBV\ with \WHI.

\acknowledgments

  The National Radio Astronomy Observatory is a facility of the National
  Science Foundation operated under contract by Associated  Universities, Inc.
  The comments of an anonymous referee are appreciated. Barry Clark, pioneer of 
  HI absorption studies, once asked the author, musingly but only partly 
  rhetorically, what HI absorption was good for, anyway.   

%%%%%%%%%%%%%%%%%%%%%

\bibliographystyle{apj}
%\bibliography{mnemonic,absorption}

%%%%%%%%%%%%%%%%%%%%

\end{document}